\documentclass[aps,twocolumn,showpacs,amsmath,amssymb,prx]{revtex4-1}

\usepackage{graphicx,tabularx}
\usepackage{bm}
\usepackage{float}
\usepackage{dcolumn}
\usepackage{color}
\usepackage{ulem}
\usepackage{mathptmx}
\usepackage{moreverb}

\renewcommand{\thefigure}{\arabic{figure}}


\begin{document}
\renewcommand{\figurename}{\textbf{Fig.}}
\renewcommand{\thefigure}{\textbf{\arabic{figure}}}

\title{
Unusual high-field metal in a Kondo insulator
}

\author{Ziji Xiang$^1$}
\email{zixiang@umich.edu}
\author{Lu Chen$^1$, Kuan-Wen Chen$^1$, Colin Tinsman$^1$, Yuki Sato$^2$, Tomoya Asaba$^{1,3}$,
Helen Lu$^4$, Yuichi Kasahara$^2$, Marcelo Jaime$^4$, Fedor Balakirev$^4$,
Fumitoshi Iga$^5$, Yuji Matsuda$^2$}
\author{John Singleton$^4$}
\email{jsingle@lanl.gov}
\author{Lu Li$^1$}
\email{luli@umich.edu}

\affiliation{
$^1$Department of Physics, University of Michigan, Ann Arbor, MI 48109, USA\\
$^2$Department of Physics, Kyoto University, Kyoto 606-8502, Japan\\
$^3$MPA-Q, Los Alamos National Laboratory, Los Alamos, NM, USA\\
$^4$National High Magnetic Field Laboratory (NHMFL), MS E536, Los Alamos National Laboratory, Los Alamos, NM 87545\\
$^5$Institute of Quantum Beam Science, Graduate School of Science and Engineering, Ibaraki University, Mito 310-8512, Japan
}
\date{\today}

\maketitle
{\bf Within condensed-matter systems, strong electronic interactions often lead to exotic quantum phases.
A recent manifestation of this is the unexpected observation of magnetic quantum oscillations~\cite{Li2014,Tan,Xiang2017,Xiang2018}
and metallic thermal transport~\cite{Hartstein,Sato2019}, both properties of systems with Fermi surfaces of itinerant quasiparticles,
in the Kondo {\it insulators} SmB$_6$ and YbB$_{12}$.
To understand these phenomena, it is informative to study their evolution
as the energy gap of the Kondo-Insulator state is closed by a large magnetic field.
We show here that both the quantum-oscillation frequency and the cyclotron mass display a strong field dependence
in the resulting high-field metallic state in YbB$_{12}$. By tracking the Fermi-surface area, we conclude that the same
quasiparticle band gives rise to the quantum oscillations in both insulating and metallic states.
These data are understood
most simply using a two-fluid picture where unusual quasiparticles, contributing little or nothing to charge transport,
coexist with conventional fermions. In the metallic state this leads to a heavy-fermion bad metal with
negligible magnetoresistance, relatively high resistivity and a very large Kadowaki-Woods ratio,
underlining the exotic nature
of the fermion ensemble inhabiting YbB$_{12}$.}

In Kondo insulators (KIs), an energy gap is opened up by strong coupling between a
lattice of localized moments and the extended electronic states.
The resulting Kondo gap $E_{\rm g}$ is usually narrow (typically $E_{\rm g} \simeq 5-20$~meV),
yet the r\^{o}le it plays in charge transport is more complicated than that of the bandgap in conventional semiconductors.
A low-temperature $(T)$ saturation of the resistivity $\rho$
has long been known in two prototypical KIs,
SmB$_6$ and YbB$_{12}$ \cite{Cooley, Kasaya}; both are mixed-valence compounds with strong
$f-d$ hybridization that defines the band structure close to the Fermi energy.
While the saturation might suggest additional metallic conduction channels, the high resistivity value within
the weakly $T$-dependent ``plateau" implies an unconventional nature for such channels~\cite{Cooley,Xiang2018}.

Recently, magnetic quantum oscillations, suggestive of a Fermi surface (FS), and thus totally unexpected in an insulator,
have been detected in both SmB$_6$ and YbB$_{12}$~\cite{Li2014, Tan, Xiang2017, Xiang2018}.
Whilst some have attributed the oscillations in SmB$_6$ to residual flux~\cite{Thomas}, the flux-free growth process of
YbB$_{12}$ (see Methods) excludes such a contribution.
The oscillations in YbB$_{12}$ are observed in both $\rho$ (the Shubnikov-de Haas, SdH, effect) and magnetization $M$ (the de Haas-van Alphen, dHvA, effect)
at applied magnetic fields $H$
where the gap is still finite.
The $T$-dependence of the oscillation amplitude follows the expectations of
Fermi-liquid theory~\cite{Xiang2018}.
Moreover, a contribution from gapless quasiparticle excitations to the heat capacity has been detected in both
KIs~\cite{Hartstein, Sato2019}. In particular, YbB$_{12}$ shows $T$-linear zero-field thermal conductivity,
a characteristic of itinerant fermions~\cite{Sato2019}.
The agreement of the FS parameters derived from the quantum oscillations,
heat capacity, and thermal conductivity suggests that the same quasiparticle band
is responsible~\cite{Sato2019}.

Despite this apparent consistency, the mystery remains: how can itinerant fermions exist in a gapped insulator
and transport heat but {\it not} charge? In response, many theoretical models entered the fray,
including topological surface states~\cite{Dzero2010}, magnetoexcitons~\cite{Knolle2017}, scalar
Majorana fermions~\cite{Erten}, emergent fractionalized quasiparticles~\cite{Chowdhury, Sodemann}
and non-Hermitian states \cite{Shen2018}. As these scenarios frequently envisage some form of exotic in-gap states,
it is potentially invaluable to observe how the properties of KIs evolve as the energy gap $E_{\rm g}$ closes.

The cubic rare-earth compound YbB$_{12}$ is an excellent platform on which to carry out such studies.
In YbB$_{12}$, the Kondo gap ($E_{\rm g}\approx 15$~meV~\cite{Okawa}) is closed by large $H$,
leading to an insulator-to-metal (I-M) transition at fields ranging from
$\mu_0 H_{\textrm{I-M}} \simeq$ 45-47\,T (${\bf H} \parallel [100]$) to 55-59\,T
(${\bf H} \parallel [110]$)~\cite{Sugiyama,TerashimaJPSJ2017,Xiang2018}.
Kondo correlation does not break down at the I-M transition, remaining strong to 60\,T and
beyond in the high-field metallic state; hence, this can be termed a Kondo metal (KM)~\cite{TerashimaPRL2018,Ohashi}.
In this study,
we apply both transport and thermodynamic measurements, including $\rho$, penetration depth, $M$,
and dilatometry, to YbB$_{12}$. By resolving quantum oscillations and tracking their $T$ and
$H$ dependence in the KM state, we trace the fate of the possible neutral
quasiparticles at fields above the gap closure and expose their interactions with
more conventional charged fermions.

In our YbB$_{12}$ samples, $M$ and magnetostriction data show that the I-M
transition occurs at $\mu_0H_{\textrm{I-M}}$ = 46.3\,T (${\bf H} \parallel [100]$);
the tiny valence increase of Yb ions at the transition suggested by magnetostriction reinforces
the KM nature of the high-field metallic state (see Methods and Extended Data Fig.\,1).
To probe the electronic structure of the KM state, a proximity-detector-oscillator (PDO)
was used (see Methods) for contactless
SdH effect studies.
The PDO technique is sensitive to the sample skin depth, providing a direct probe of changes in the conductivity of the
metallic KM state. The setup illustrated in Fig.\,1a (inset) was rotated on a cryogenic goniometer
to achieve {\bf H}-orientation-dependent measurements.
Fig.\,1a summarizes the $H$ dependence of the PDO frequency $f$ as ${\bf H}$ rotates
from [100] to [110].

The low conductivity of the KI state suggests that the MHz oscillatory field from the PDO coil will completely
penetrate the sample \cite{PDO_Magnetization}. Therefore, the response of the PDO is
dominated by the sample skin depth only when the sample enters the KM state and the conductivity
increases significantly (see Methods).
The transition between these regimes is marked by a dip
in the $f$ versus $H$ curves close to $H_{\rm I-M}$, above which magnetic quantum oscillations emerge.
Fig.\,1b displays $\Delta f$, the oscillatory component of $f$, at various angles $\theta$ as a
function of $1/H$.
The distinct oscillation pattern observed for ${\bf H} || [100]$ $(\theta=0)$ is preserved up to
$\theta \approx 21^\circ$ (Fig.\,1b), being strongly modified at higher angles (Extended Data Fig.\,2a).

The oscillations in $\Delta f$ represent a single series that is intrinsically aperiodic in $1/H$;
attempts to fit them using a superposition of conventional oscillation frequencies fail to
reproduce the raw data (Supplementary Information).
To demonstrate the point further, Fig.\,1c compares Landau-level indexing plots for
the low-field KI and high-field KM states, both with ${\bf H} \parallel$[100].
The oscillatory component of the
resistivity $\Delta \rho$ in the KI state is shown in the inset of Fig.~1c and indexed conventionally using integers
for minima and half integers for maxima in $\rho$.
For the KM state, peaks in $f$ correspond to peaks in conductivity~\cite{PDO_Magnetization},
and are therefore indexed using integers~\cite{Shoenberg}.
A further subdivision of the oscillations, reminiscent of a second harmonic,
is likely due to Zeeman splitting of the quasiparticle levels~\cite{Shoenberg};
these features are marked with``+" and ``-" assuming signs expected for conventional Zeeman shifts.
In the KI state, the plot of Landau level index $N$ versus $1/H$ is a straight line, as expected
for a field-independent quantum oscillation frequency in a nonmagnetic system.
By contrast, in the KM state the $1/H$ positions of the oscillations have a nonlinear
relationship with $N$ which we shall describe below.
(Here we note that the magnetic induction $B \approx \mu_0 H$, since $\mu_0H$ is large and YbB$_{12}$ has a weak
magnetization; see Methods).

Nevertheless, despite their unusual periodicity, the $T$-dependences of individual oscillation amplitudes
in the KM state (Fig.\,1d)
closely follow the Lifshitz-Kosevich (LK) equation~\cite{Shoenberg},
\begin{equation}
\Delta f(T) \propto \frac{2\pi^2 k_{\rm B} T/E_{\rm c^*}}{\sinh(2\pi^2 k_{\rm B} T/E_{\rm c^*})},
\label{LK_temp}
\end{equation}
suggesting that they are almost certainly due to fermions.
(Here, $k_{\rm B}$ is the Boltzmann constant and $E_{\rm c^*}$ the cyclotron energy).
However, the derived value of $E_{\rm c^*}$ varies nonlinearly with $H$, indicating that
the cyclotron mass $m^* = eB/E_{\rm c^*}$ is a function of $H$ (Fig.\,2b, inset).

The relationship between oscillation indices and magnetic field is described empirically by
\begin{equation}
N + \lambda = \frac{F_0}{\mu_0(H_N-H^*)},
\label{LinearLL}
\end{equation}
where an offset field of $\mu_0 H^* = 41.6$~T has been subtracted from $\mu_0H_N$.
Here, $\lambda$ is a phase factor, $N$ is again the index, $H_N$ is the
field at which the corresponding feature ({\it e.g.,} peak) occurs and $F_0$ is the slope.
Eq.~\ref{LinearLL} is symptomatic of a FS
pocket that progressively depopulates as $H$ increases, for reasons that
we will discuss below.
In such cases, the Onsager relationship
$F(B)=\frac{\hbar}{2\pi e} A(B)$ between the
FS extremal cross-sectional area $A$ and the frequency $F$ of the corresponding
quantum oscillations still applies even when $A$ changes with $H$ (see Methods).
Since $B\approx \mu_0 H$, we write $B^*=\mu_0H^*$
and represent the field dependence of Eq.~\ref{LinearLL} using a $B$-dependent frequency
\begin{equation}
F_{\rm KM}(B)= \frac{F_0}{B-B^*}B;
\label{OrbitF}
\end{equation}
this is associated with a $B$-dependent extremal area
$A(B)= \frac{A_0}{B-B^*}B$,
where $A_0$ = $\frac{2\pi e}{\hbar}F_0$ \cite{backproj}.

Analyzed in these terms, our data indicate that the quantum oscillations seen in the KM phase
are due to a FS pocket that is the same as, or very closely related to, the FS pocket
in the KI state which contributes
quantum oscillations and a $T$-linear term in the
thermal conductivity and heat capacity~\cite{Sato2019}.
This is strongly suggested by the field dependence of $F_{\rm KM}$ (Eq.~\ref{OrbitF}) shown in Fig.\,2b.
Even a cursory inspection reveals that Eq.~\ref{OrbitF},
describing oscillations in the KM
state, gives a frequency very similar to that of
the KI-state oscillations when extrapolated back to $H_{\rm I-M}$.
This is true for all angles $\theta$ at which oscillations
were measured; using appropriate $F_0$ values (Extended Data Table\,1)
and substituting
fields $B =\mu_0 H_{\rm I-M}(\theta)$ on the phase boundary (minima in PDO data; Fig.~1a)
into Eq.~\ref{OrbitF}
gives the frequencies shown as magenta diamonds in Fig.\,2c.
The $\theta-$dependence of $F_{\rm KM}(\mu_0 H_{\rm I-M})$
thus deduced tracks the
behaviour of the dHvA frequencies measured in the KI state,
albeit with an offset $\approx 100$~T.
This offset may be due to a discontinuous change in $F$ at the phase boundary;
however, it could also result from the potential uncertainty
in determining the $H$-position of any phase transition associated with a valence change~\cite{Ho}.
Substituting values of $\mu_0H_{\rm I-M}$ decreased by $\approx 0.8$~T
into Eq.~\ref{OrbitF} yields an exact match of $F_{\rm KM}$ with the
quantum-oscillation frequencies observed in the KI state (Fig.\,2c, red diamonds).

This consistency indicates that
the novel quasiparticles detected in the KI state of YbB$_{12}$ \cite{Xiang2018},
which are probably charge neutral~\cite{Sato2019}, also cause the SdH effect in the KM state
(see Methods for further discussion).
The unusual nature of the KM state is further revealed by magnetotransport experiments.
To reduce Joule heating as the KI state is traversed, we used a pulsed-current technique (see Methods).
Current is only applied to the sample when $H>H_{\textrm{I-M}}$,
as sketched in Fig.\,3a, inset.
Below 10\,K, very weak longitudinal magnetoresistance (MR) is observed above 50\,T (Extended Data Fig.\,3)
and is preserved up to $\approx 68$~T
(Extended Data Fig.\,2), permitting an analysis of the $T$-dependence of the resistivity $\rho$ in the KM state.
Fig.\,3a and Fig.\,3b show $\rho$ at 55\,T as a function of $T$ and $T^2$, respectively.
The $\rho-T$ curve shows a maximum at $T^*$ = 14~K. With decreasing $T$, a linear $T$-dependence develops
below 9\,K and extends down to $T \approx 4$~K; subsequently, a Fermi-liquid-like $T^2$-behavior is established below
$T_{\textrm{FL}}$ = 2.2\,K. The overall behavior of $\rho$($T$) mimics that of a typical Kondo lattice
where the Kondo coherence develops at $T^*$ and a heavy Fermi-liquid state forms below
$T_{\textrm{FL}}$~\cite{YFYang}. The residual resistivity $\rho(T\rightarrow0)\sim$ 0.4 m$\Omega\cdot$cm indicates that the KM state
may be classified as a ``bad metal".

The negligible MR indicates that {\it even if} the FS pocket responsible for the quantum
oscillations is capable of carrying charge in the KM state, it plays a negligible r\^{o}le in $\rho$ (see Methods).
Moreover, this FS pocket would only account for 4-5$\%$ of the Sommerfeld coefficient obtained from
 pulsed-field heat-capacity measurements ($\gamma \simeq 63$~mJ mol$^{-1}$ K$^{-2}$ at
 55\,T~\cite{TerashimaPRL2018}, see Methods). Therefore, a separate FS of more conventional heavy fermions
 is {\it required} to account for the charge-transport properties and the large Sommerfeld coefficient of the KM state.
 However, the heavy effective mass of these fermions (see Methods) and the relatively
 high $T > 0.4$~K used for our measurements will preclude observation of their quantum oscillations.

In view of the relatively high resistivity in the KM state, the Kadowaki-Woods (KW) ratio $A_2$/$\gamma^2$ ($A_2$ is
the $T^2$-coefficient of resistivity) is surprisingly large. Using the value of $A_2$  given by the fit in Fig.\,3b,
the estimated KW ratio is $1.54 \times 10^{-2}~\mu\Omega$cm (K mol/mJ)$^2$, three and four orders of
magnitude larger than typical values for heavy-fermion compounds and transition metals, respectively (Fig.\,3c).
Such an abnormal KW ratio cannot be addressed by the degree of degeneracy of quasiparticles, which tends to
suppress the KW ratio in many Yb-based systems~\cite{Tsujii, Matsumoto}.

The data in Figs.~1-3 imply an intriguing two-fluid picture in YbB$_{12}$ that includes
(i)~a FS pocket of quasiparticles obeying Fermi-Dirac statistics but contributing little to the transport of charge;
and (ii)~more conventional charged fermions. For brevity, we refer to (i) as neutral fermions (NFs).
These NFs cause quantum oscillations in both KI and KM states, whereas (ii) dominate the
electrical transport properties and the low-$T$ heat capacity in the KM state.
Under this two-fluid description, the I-M transition produces a sudden increase of the density of (ii),
which changes from a thermally excited low-density electron gas in the KI state to a dense liquid of heavy,
charged quasiparticles in the KM state. This change has two consequences.
First, it dramatically enhances the charge screening; this will in turn affect the interactions that contribute to the
renormalization of the effective mass of the NFs~\cite{quader}.
This is the likely cause of the significant fall in mass as $H$ increases through the I-M transition
that is suggested if one extrapolates the $m^*$ versus $H$ plot (Fig.~2b, inset) back to the phase boundary.
Second, on entering the KM state, the much enhanced number of states available close to the Fermi energy will
act as a ``reservoir'' into which quasiparticles from the NF FS can scatter or transfer (for analogous
situations in other materials, see~\cite{khg,hgcdte}). As $H$ increases past the I-M transition,
it appears that the NF FS becomes less energetically favourable; the availability of the KM ``reservoir''
means that quasiparticles can transfer out, leading to the falling quantum-oscillation frequency
parameterized by Eq.~\ref{OrbitF}. As the NF FS shrinks, the effective mass increases (Fig.~2b),
possibly due to nonparabolicity of the corresponding band and/or field-induced modification of
the bandwidth or interactions contributing to the effective mass renormalization~\cite{quader}.
Also, the fact that the quantum oscillations caused by the NF FS are strongly affected by the I-M transition ({\it i.e.,}
the frequency becomes $H$ dependent in the KM phase) shows that they are an intrinsic property of YbB$_{12}$,
laying to rest suggestions that they are caused by a minority phase ({\it c.f.} the Al flux
proposal for SmB$_6$~\cite{Thomas}) or surface effect.

The survival of the NFs above the gap closure and their coexistence with charged excitations in a Fermi-liquid-like
state \cite{Chowdhury,Senthil2003,Senthil2004} leads to an interesting scenario; in such a state, Luttinger's theorem may be violated and a continuous
variation of FS properties is allowed~\cite{Senthil2004}. In addition, the interaction of a relatively conventional
FS with an ensemble of quasiparticles unable to transport charge is a feasible prerequisite for $\rho$ varying
linearly with $T$ (Fig.~3a)~\cite{MarginalFL}. Finally, the scaling behaviour between the magnetization and the
$B$-dependence of orbit area (see Supplementary Information), as well as the
large ``offset" field $H^*$ needed to linearize the Landau diagrams (Fig.\,2a), may suggest underlying nontrivial
magnetic properties~\cite{gaugefield} in the KM state that await further investigation.

\section*{Data availability}
The data that support the plots within this paper and other findings of this study are available from the corresponding authors upon reasonable request.

\section*{Code availability}
The code that support the plots within this paper and other findings of this study are available from the corresponding authors upon reasonable request.

\section*{Acknowledgements}

We thank Lin Jiao, Liang Fu, Senthil Todadri, Tom Lancaster, Paul Goddard, Hiroshi Kontani, Hiroaki Shishido
and Robert Peters for discussions.
This work is supported by the National Science Foundation under Award No. DMR-1707620
(electrical transport measurements), by the Department of Energy under Award No. DE-SC0020184
(magnetization measurements), by the Office of Naval Research through DURIP Award
No.~N00014-17-1-2357 (instrumentation), and  by Grants-in-Aid for Scientific Research (KAKENHI) (Nos. JP15H02106, JP18H01177, JP18H01178, JP18H01180, JP18H05227, JP19H00649, JP20H02600 and JP20H05159) and on Innovative Areas ``Quantum Liquid Crystals" (No. JP19H05824) from Japan Society for the Promotion of Science (JSPS), and JST CREST (JPMJCR19T5).
A major portion of this work was performed at the National High Magnetic Field Laboratory, which is supported by
National Science Foundation Cooperative Agreement No. DMR-1644779 and the Department of Energy (DOE).
J.S. and M.J. thank the DOE for support from the BES program ``Science in 100~T".
The experiment in NHMFL is funded in part by a QuantEmX grant from ICAM and the
Gordon and Betty Moore Foundation through Grant GBMF5305 to Dr. Ziji Xiang, Tomoya Asaba,
Lu Chen, Colin Tinsman, and Dr. Lu Li.
We are grateful for the assistance of You Lai, Doan Nguyen, Xiaxin Ding, Vivien Zapf, Laurel Winter, Ross McDonald and Jonathan Betts of the NHMFL.

\section*{Author contributions}
F.I. grew the high-quality single crystalline samples. Z.X., L.C., K-W.C., C.T., Y.S., T.A., H.L., F.B., J.S. and L.L.
performed the pulsed field PDO and resistivity measurements. Z.X., T.A. and J.S. performed the pulsed field magnetometry
measurements. L.C., C.T. and M.J. performed the pulsed field magnetostriction measurements.
Z.X., K-W.C., Y.K., Y.M., J.S. and L.L. analyzed the data. Z.X., Y.M., J.S. and L.L. prepared the manuscript.

\section*{Author information}
The current address of M.J. is Physikalisch-Technische Bundesanstalt, Braunschweig 38116, Germany. The authors declare no competing financial
interests. Correspondence and requests for materials should be addressed to Z.X.(zixiang@umich.edu), J.S.(jsingle@lanl.gov) and L.L.(luli@umich.edu).

\newpage
{\bf Methods}\\
\noindent
{\bf Sample preparation and pulsed field facilities.}
YbB$_{12}$ single crystals were grown by the traveling-solvent floating-zone method \cite{Iga}.
The two samples studied in this work were cut from the same ingot and shown to have almost identical physical
properties, including the SdH oscillations below the I-M transition, in our previous investigations~\cite{Xiang2018,Sato2019}.
The YbB$_{12}$ sample characterized in the magnetostriction and the magnetization ($M$)
measurements corresponds to
sample N1 in Ref.\,\cite{Xiang2018} and \#2 in Ref.\,\cite{Sato2019}, whereas the high-field
MR was measured in the
YbB$_{12}$ sample N3 in Ref.\,\cite{Xiang2018}, which is also crystal \#1 in Ref.\,\cite{Sato2019}.
Both samples were used in the PDO experiments.
The PDO data from N3 taken at a fixed field direction were published elsewhere \cite{Xiang2018}.

Magnetostriction and $M$ of YbB$_{12}$ samples were measured in a
capacitor-driven 65\,T pulsed magnet at NHMFL, Los Alamos.
In the PDO and MR measurements, fields were provided by 65\,T
pulsed magnets and a 75\,T Duplex magnet.
Temperatures down to 500~mK are obtained using a $^3$He immersion cryostat.
In the MR experiment, the sample was immersed in liquid $^4$He to achieve a more precise
$T$ in the range $1.3\leq T \leq 4.0$~K.

\noindent
{\bf Magnetostriction measurements.}
The linear magnetostriction $\Delta L$/$L$ of YbB$_{12}$ was measured using a
fibre Bragg grating dilatometry technique~\cite{Daou,Jaime}.
In our setup (Extended Data Fig.\,1), the dilatometer is a 2~mm-long Bragg grating contained in a $125~\mu$m
telecom-type optical fibre. The oriented YbB$_{12}$ single crystal was attached to the section of fibre with the Bragg grating
using a cyanoacrylate adhesive.
The crystallographic [100] direction was aligned with the fibre, which is also parallel to $H$.
Thus, we measure the longitudinal magnetostriction along the $a$-axis of cubic YbB$_{12}$.
The magnetostriction $\Delta L$/$L$ was extracted from the shift of the Bragg wavelength in the reflection spectrum~\cite{Jaime}.
The signal from an identical Bragg grating on the same fibre with no sample attached was subtracted as the background.

In a paramagnetic metal, the high-field longitudinal magnetostriction contains both $M^2$ and $M^3$ terms~\cite{YbInCu4}.
In this sense, the power-law $H$ dependence of
$\Delta L/L$ with an exponent $\approx 3.5$ (Extended Data Fig.\,1b) is consistent with the
weak superlinear $M$ in YbB$_{12}$ at 40\,K~\cite{Sugiyama}.
As $T$ lowers, $\Delta L$/$L$ decreases and a nonmonotonic field dependence develops at 30\,K (Extended Data Fig.\,1b).
We note that the fast suppression of $\Delta L/L$ coincides with the sharpening of the I-M transition in the
derived susceptibility below 30\,K~\cite{YHMatsuda}, suggesting an additional energy scale in YbB$_{12}$
that is much lower than the Kondo temperature $T_{\rm K} \approx 240$~K and the gap opening temperature
$T_{\rm g} \approx 110$~K \cite{Okawa}. Below 5\,K, $\Delta L/L$ becomes quite small and a step-like feature
is observed with an onset at $\mu_0 H$ = 46.3\,T, perfectly aligned with the sudden increase in $M$ (Extended Data Fig.\,1c,d).
We identify this characteristic field as the I-M transition field $H_{\textrm{I-M}}$ at
which a metamagnetic transition also happens~\cite{TerashimaJPSJ2017,TerashimaPRL2018,YHMatsuda}.
A negative volume magnetostriction is characteristic of mixed-valence Yb compounds in which
the volume of nonmagnetic Yb$^{2+}$ (4$f^{14}$) is 4.6$\%$ smaller than that of magnetic Yb$^{3+}$ (4$f^{13}$).
Therefore a volume decrease with increasing $H$ is expected \cite{Yoshimura,Volume4f}.
The step-like decrease at $H_{\textrm{I-M}}$ (Extended Data Fig.\,1c)
may therefore be evidence that the sudden shrinkage results from a valence transition of the Yb ions.
Using a simple isotropic assumption, the change of volume magnetostriction at $H_{\textrm{I-M}}$ is
$\delta(\Delta V/V) = 3\delta(\Delta L/L) \simeq 6 \times 10^{-6}$, corresponding to a
valence increase of 0.00013. Such a small average valence enhancement implies a quite weak
and incomplete breakdown of the Kondo screening. Consequently, the state immediately
above $H_{\textrm{I-M}}$ is confirmed to be a KM in which mixed-valence features persist.

\noindent
{\bf Magnetization measurements.}
$M$ was measured using a compensated-coil susceptometer~\cite{GoddardPRB2007,GoddardNJP2008}.
The 1.5~mm bore, 1.5~mm long, 1500-turn coil is made of 50~gauge high-purity copper wire.
The sample was inserted into a 1.3~mm diameter non-magnetic ampoule that can be moved in and out of the coil.
When pulsed fields are applied, the coil picks up a voltage signal $V \propto ({\rm d}M/{\rm d}t)$, where $t$ is the time.
Numerical integration is used to obtain $M$ and a signal from the empty coil measured under identical conditions is subtracted.
Pulsed-field $M$ data are calibrated using the $M$ of a YbB$_{12}$ sample of known mass
measured in a Quantum Design VSM magnetometer~\cite{Sato2019}.

As shown in Extended Data Fig.\,1d, a metamagnetic transition occurs at 46.3\,T, coinciding
with the onset of the step-like feature in the magnetostriction.
This observation further confirms the location of $H_{\textrm{I-M}}$ in our YbB$_{12}$ samples.
At the highest $H$ used in this experiment, $M\approx 1 \mu_{\rm B}$/Yb,
so that $M$ contributes only $\sim$ 0.2$\%$ of $B$.
Therefore we can ignore the $M$ term and equate $B$ to the external magnetic field, {\it i.e.,} $B \approx \mu_0 H$.

\noindent
{\bf Radio frequency measurements of resistivity using the PDO technique.}
The PDO circuit~\cite{PDO_Moaz,PDO_Magnetization} permits convenient contactless measurements of the resistivity
of metallic samples in pulsed magnetic fields. In our experiments, a 6-8~turn coil made from 46~gauge high-purity
copper wire is tightly wrapped around the YbB$_{12}$ single crystals and secured using GE varnish.
The coil is connected to the PDO,  forming a driven LC tank circuit with a resonant
frequency of 22-30~MHz at cryogenic $T$ and $H=0$. The output signal is fed to a two-stage mixer/filter
heterodyne detection system~\cite{PDO_Magnetization}, with mixer IFs  provided by a dual-channel BK-Precision
Function/Arbitrary Waveform Generator. The second mixer IF was 8~MHz, whereas the
first mixer IF was adjusted to bring the final frequency down to $\approx 2$~MHz.
The resulting signal was digitized using a National Instruments PXI-5105 digitizer.

Considering all the contributions, the shift in PDO frequency $f$ due to $H$ is
written as \cite{Xiang2018,PDO_Magnetization}
\begin{equation}
\Delta f = -a \Delta L - b \Delta R,
\label{PDO}
\end{equation}
where $a$ and $b$ are positive constants determined by the frequency plus the capacitances,
resistances and inductances in the circuit, $L$ is the coil inductance and $R$ is the resistance of the coil wire and cables.
In the case of a metallic sample, the coil inductance $L$ depends on the skin depth $\lambda$ of the sample.
If we assume that the sample magnetic permeability $\mu$ and the coil length stay unchanged during a field pulse,
we have $\Delta L \propto (r-\lambda)\Delta \lambda$, where $r$ is the sample radius.
At angular frequency $\omega$, the skin depth is proportional to square root of the resistivity $\rho$:
\begin{equation}
\lambda = \sqrt{\frac{2 \rho}{\omega \mu}}.
\end{equation}
Therefore, for a metallic sample, the resonance shift $\Delta f$ reflects the sample MR
and the detected quantum oscillations are due to the SdH effect.
In YbB$_{12}$, the PDO measurement only detects the signal from the sample in the high-field KM state,
{\it i.e.,} when $H>H_{\textrm{I-M}}$ \cite{Xiang2018}. In the low-field KI state the sample is so resistive that
the skin depth $\lambda$ is larger than the sample radius $r$. As a result, $\Delta f$ mainly comes from
the MR of the copper coil~\cite{PDO_Magnetization}. The ``dip" in PDO $f$ in Fig.\,1a
consequently indicates where the skin depth is comparable to the sample radius and provides an alternative
means to find $H_{\rm I-M}$. We note that the $H_{\textrm{I-M}}$ assigned to onset of the ``dip" feature
(see Extended Data Table\,1) is $\approx 0.2$~T lower than the metamagnetic transition in Extended Data Fig.\,1.

\noindent
{\bf Onsager relationship for a field-dependent Fermi surface.}
The Onsager relation~\cite{Shoenberg} relates the frequency $F$ of quantum oscillations to the
FS extremal orbit area $A$: $F=\frac{\hbar}{2\pi e} A$.
Textbook derivations~\cite{Shoenberg} invoke the Correspondence Principle to give
an orbit-area quantization condition $(N+\lambda)\frac{2\pi e B}{\hbar} = A$, where $N$ is a quantum number  and $\lambda$ is a phase factor.
The derivation makes no assumptions about $A$ being constant, so for a field-dependent  $A=A(B)$,
we can write
\begin{equation}
(N+\lambda)\frac{2\pi e B_N}{\hbar} = A(B_N),
\label{quantized_area}
\end{equation}
where $B_N$ is the magnetic induction at
which the $N^{\rm th}$ oscillation feature (peak, valley,{\it etc.}) occurs.
Evaluating Eq.~\ref{quantized_area} for $N$ and $N+1$ and taking the difference gives
\begin{equation}
\frac{A(B_{N+1})}{B_{N+1}} - \frac{A(B_N)}{B_N} = \frac{2\pi e}{\hbar}.
\label{area_diff}
\end{equation}
From an experimental standpoint, in materials where $F$ is constant, a particular feature of a quantum oscillation is observed whenever
$N+\lambda'=\frac{F}{B}$; $\lambda'$ is another phase factor. Allowing $F$ to vary ({\it i.e.,}
$F=F(B)$), evaluating the expression for $N$ and $N+1$ and taking the difference
yields
\begin{equation}
\frac{F(B_{N+1})}{B_{N+1}} - \frac{F(B_N)}{B_N} = 1.
\label{freq_diff}
\end{equation}
A comparison of Eqs.\,\ref{area_diff} and~\ref{freq_diff} shows that the Onsager relation still holds, {\it i.e.,}  $F(B)=\frac{\hbar}{2\pi e} A(B)$.

In Fig.~2a, $B^*$ = $\mu_0 H^*$ = 41.6\,T is subtracted from the applied field to yield linear
Landau-index diagrams for $\theta \leq$ 20.7$^\circ$. The resulting fit is described by
$N+\lambda = \frac{F_0}{(B_N-B^*)}$ (Eq.~\ref{LinearLL}), which can be written in terms of a $B$-dependent frequency
$N+\lambda = \frac{F(B_N)}{B_N}$, where $F(B) = \frac{F_0 B}{(B-B^*)}$ (Eq.~\ref{OrbitF}).
Following the reasoning given above, this is associated with a $B$-dependent extremal area
\begin{equation}
A(B)= \frac{A_0}{B-B^*}B,~~~{\rm with}~~~ A_0 = \frac{2\pi e}{\hbar}F_0.
\label{Orbit}
\end{equation}

In the Supplementary Information, we show that the analysis presented here is functionally
equivalent to the ``back-projection'' method~\cite{backproj}.
Note that we have not considered the Zeeman splitting of the peaks, so that
Eqs.\,\ref{OrbitF} and\,\ref{Orbit} describe the average of the spin-up and spin-down components.

\noindent
{\bf Resistivity measurements in the KM state.}
From $\mu_0H=0$ to 60\,T, the resistivity of YbB$_{12}$ decreases by five orders of magnitude~\cite{Xiang2018}.
As a result, MR measurements in pulsed fields are challenging.
If a constant current is used during the entire field pulse, either the signal-to-noise ratio is poor in the
KM state, or the high resistance in the KI state causes serious Joule heating.
To solve this issue, we developed a pulsed-current technique.
The experimental setup is shown in Extended Data Fig.\,3a.
A ZFSWHA-1-20 isolation switch is used to apply current pulses with widths $<10$~ms.
The switch is controlled by two square wave pulses generated by a BK-Precision Model 4065 dual-channel
function/arbitrary waveform generator triggered by the magnet pulse.
Thus, a relatively large electric current (2-3~mA) can pass through the sample during
a narrow time window within the field pulse (Fig.\,3a, inset).

Current is applied only when YbB$_{12}$ enters the KM state. In our low-$T$ MR measurements,
the switch turns on at 47\,T during the upsweeps and turns off at 47\,T during the downsweeps.
To reduce heating due to eddy currents, we measured the longitudinal MR of a needle-shaped
sample (length 6.5~mm between voltage leads; cross-sectional area 0.33~mm$^2$).
As shown in Extended Data Fig.\,3b, the downsweeps still suffer some Joule heating.
Hence, in the main text we focus on the upsweeps, which show very weak longitudinal MR and
should reflect the intrinsic electrical transport properties of the KM state of YbB$_{12}$.
To further dissipate heat, we measured the pulsed field MR with the sample in liquid $^4$He
for $1.3 \leq T \leq 4.0$~K;
in this range, $^3$He sometimes gives a poor thermal impedance on the sample surface.
For this reason, the linear fit in Fig.\,3b only uses data with the sample in liquid $^4$He.

According to Eq.\,\ref{Orbit}, $A(B)$ for the pocket detected in the PDO measurement shrinks by $\simeq$ 45$\%$
from 50\,T to 60\,T. Assuming a spherical FS, this corresponds to an $\simeq 60\%$ reduction in
quasiparticle density $n$. Meanwhile, the cyclotron mass increases by $\simeq 60\%$ (Fig.\,2b, inset).
Consequently, a textbook Drude expression $\rho = m^*/ne\tau$ ($\tau$ is the relaxation time) predicts that the resistivity
would increase by a factor $\approx 4$ from 50\,T to 60\,T if the electrical transport is dominated by this pocket.
In sharp contrast, almost no MR is observed in this field range (Extended Data Fig.\,3), indicating
the negligible contribution of this FS pocket to the electrical transport.

On the other hand, the weak MR and the $T$-linear resistivity of the KM state
between 4\,K and 9\,K are
in agreement with the predicted behaviour of a marginal Fermi liquid (without macroscopic disorder)
or an incoherent metal, potential evidence for the coexistence of fermions inactive in the charge transport
and more conventional charged fermions~\cite{MarginalFL} .
\bigskip

\noindent
{\bf The Kadowaki-Woods ratio.} For the Sommerfeld coefficient $\gamma$ in the KM state of YbB$_{12}$,
a pulsed-field heat capacity study reports values of 58 mJ mol$^{-1}$ K$^{-2}$ and 67 mJ mol$^{-1}$ K$^{-2}$ at 49\,T
and 60\,T, respectively \cite{TerashimaPRL2018}. A linear interpolation gives $\gamma$ = 63 mJ mol$^{-1}$ K$^{-2}$ at 55\,T.
Since $\gamma$ = ($\pi^2 k_\textrm{B}^2$/3)$N$($E_\textrm{F}$) and the density of states at the Fermi
 energy $N$($E_\textrm{F}$) = ($m^*/\pi^2 \hbar^2$)(3$\pi^2 n$)$^{1/3}$, the Sommerfeld coefficient can be written as:
\begin{equation}
\gamma = \frac{\pi^2 k_B^2}{3} \frac{m^* k_{\rm F}}{\pi^2 \hbar^2},
\label{gamma}
\end{equation}
where $k_{\rm F}$ is the Fermi vector. As for the FS pocket detected in the SdH measurement,
Eq.\,\ref{OrbitF} gives $F$(55\,T) = 231.9\,T. In a spherical FS model,
$F$ = $\hbar k_F^2/2 e$, $n$ = $k_F^3$/3$\pi^2$ = $(2eF/\hbar)^{3/2}$/3$\pi^2$
($k_{\rm F}$ is the Fermi wavevector), therefore $F$ = 231.9\,T corresponds to $n$ = 1.99$\times$10$^{19}$ cm$^{-3}$.
Also, for 55~T, $m^* \simeq 10 m_{\rm e}$, based on Fig.\,2b (inset).
Putting these parameters into Eq.\,\ref{gamma}, we estimate that the FS pocket responsible for the SdH effect
could contribute only 4.4$\%$ of the measured $\gamma$. Consequently, an additional band with a
much larger density of states must exist in the KM state of YbB$_{12}$.

The unusually large KW ratio (Fig.\,3c) suggests the somewhat unusual nature of the
heavy quasiparticles that dominate the charge transport and thermal properties of the KM state.
Analysis of the KW ratio shows that for a single-band, strongly correlated system,
the value of $A_2$/$\gamma^2$ does not depend on the strength of correlations
but is instead solely determined by the underlying band structure~\cite{Jacko}.
We consider the simplest case of a single-band isotropic Fermi-liquid model.
Using the same calculations as in Ref.\,\cite{Jacko}, we have:
\begin{equation}
A_2 = \frac{81\pi^3 k_B^2}{4e^2 \hbar^3} \frac{(m^*)^2}{k_F^5}.
\label{A2}
\end{equation}
Taken together, Eqs.\,\ref{gamma} and~\ref{A2} yield $k_{\rm F} = 2.126~{\rm nm}^{-1}$
(corresponding to $n = 3.25 \times 10^{20}~{\rm cm}^{-3}$)
and a rather large effective mass of $m^* = 89.6 m_{\rm e}$.
Such a heavy mass is unusual for Yb-based mixed-valence compounds,
but explains the anomalous KW ratio as well as the absence of
SdH oscillations due to these quasiparticles in the PDO response.

\bigskip

\noindent
{\bf SdH oscillations due to charge-neutral quasiparticles.} Magnetic quantum
oscillations are observed in both $M$ and $\rho$ in the KI phase of
YbB$_{12}$ \cite{Xiang2018}. The size of the FS and the effective mass of the quasiparticles
inferred from the oscillations are consistent with the fermion-like contribution to the
thermal conductivity \cite{Sato2019}. A natural explanation, given the high electrical resistivity,
is that the thermal conductivity and quantum oscillations are due to charge-neutral fermions.

Of the two oscillatory effects observed in the KI phase, the dHvA effect is the more fundamental.
As pointed out by Lifshitz, Landau and others \cite{Shoenberg} , it involves oscillations in a
thermodynamic function of state - $M$ - that may be related to the electronic density of states
with a minimum of assumptions.
The fact that $M$ in the KI phase oscillates as a function of $H$ can only be due to
the oscillation of the fermion density of states, and consequently their free energy.

Even in conventional metals, quantum oscillations in $\rho$ are
harder to model quantitatively.
A starting point was suggested by Pippard~\cite{Shoenberg}; the rate at which
quasiparticles scatter will depend on the density of states available via {\it Fermi's Golden Rule}.
Hence, if the quasiparticle density of states oscillates as a function of $H$, $\rho$ will also
oscillate proportionally, leading to the SdH effect.

Before modifying this idea to tackle the SdH effect in the KI phase, we remark that the
$H$-dependent frequency of the oscillations in the KM phase suggests that exchange
of fermions between charge-neutral and conventional FS sections occurs readily,
probably via low-energy scattering. This is also supported by the $T$-linear resistivity \cite{MarginalFL}.
The rate at which this scattering occurs will obviously reflect the joint density of fermion states.

Returning to the KI phase of YbB$_{12}$, $\rho$ is thought to be due to charge carriers
thermally excited across the energy gap, plus contributions from
states in the gap that lead to the $\rho$ saturation at low $T$.
Following the precedent of the KM phase, it is likely that fermions in the KI phase scatter back and forth
between the charge-neutral states and the more conventional bands.
The situation is more complicated than Pippard's original concept because scattering {\it between} two bands
is involved. Nevertheless, the amount of scattering will be determined by the joint density of states,
and, because the density of states of the neutral quasiparticles oscillates with $H$, so will $\rho$.
Unlike conventional metals, the density of the charge carriers will be $T$-dependent, so the amplitude of the
SdH oscillations in the KI phase will follow a different $T$dependence from that of the dHvA oscillations, as observed in experiments~\cite{Xiang2018,SodMan}.

As described by Eq.\,\ref{PDO}, the PDO $f$ in the KM phase is determined by the skin depth,
{\it i.e.,} the conductivity. How then, can oscillations due to the neutral fermions be manifested
in this signal? As described above, fermions will scatter back and forth between the charge-neutral states
and more conventional bands; in the KM phase, our conductivity results and the heat capacity
data of others suggests that the latter is a  FS of heavy charged fermions.
Once again, scattering between two bands will occur, and the SdH effect in conductivity caused by the oscillatory
density of states of the neutral quasiparticles in magnetic fields is detected in the PDO experiments.
By contrast, any intrinsic quantum oscillations due to the metallic FS will be suppressed by a combination of
the very heavy effective mass and the relatively high  $T~(\gtrsim 0.4$\,K) of our measurements.

\newpage

\begin{figure}[hbtp]
	\begin{center}
		\includegraphics[width=\linewidth]{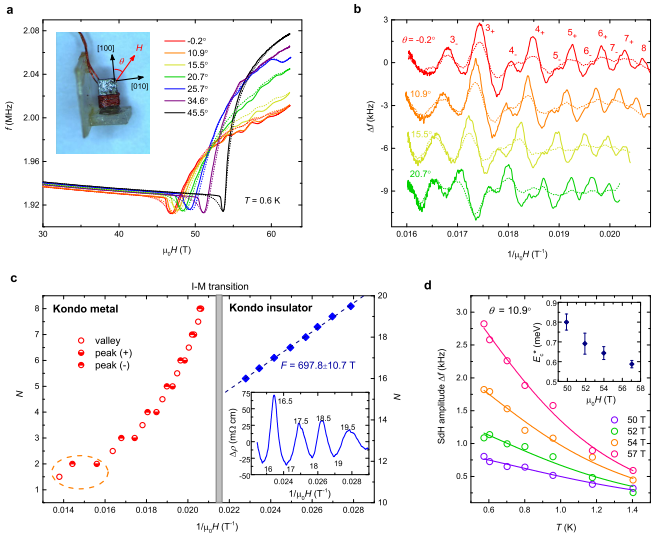}
		\caption{{\bf The SdH effect in YbB$_{12}$.} {\bf a,} The PDO frequency $f$ for a YbB$_{12}$ single-crystal sample,
		measured at $T$ = 0.6\,K at various tilt angles $\theta$ up to 62.5\,T.
		The ``dip" feature in $f$ corresponds to the I-M transition which shifts to higher field at larger tilt angles.
		Solid lines and short-dashed lines are upsweeps and downsweeps, respectively.
		Inset: a photograph of the device we used in pulsed-field PDO experiments. The whole device was attached to a rotation
		stage with the rotation axis normal to the (001)-plane. The tilt angle $\theta$ is defined as the angle between the field vector ${\bf H}$
		and the crystallographic [100] direction.
		{\bf b,} Oscillatory component of the PDO frequency, $\Delta f$, obtained after a 4th-order polynomial background subtraction
		from the raw data shown in {\bf a} for different tilt angles from $\theta = 0.2^\circ$ to $\theta = 20.7^\circ$.
		The numbers beside the SdH peaks are the Landau level indices.
		The signs ``+" and ``-" mark the spin-split Landau sublevels with inferred Zeeman shift +$g\mu_B B$/2 (spin-down)
		and -$g\mu_B B$/2 (spin-up),
		respectively. The SdH effect is weaker on the downsweeps (short-dashed lines), probably due to sample heating.
		{\bf c,} Landau-level plots for YbB$_{12}$ in the low-field KI state (blue diamonds) and the high-field KM state
		(red circles), both under a magnetic field along the [100] direction. Data points in the orange dashed circle were taken
		in the 75\,T Duplex magnet (Extended Data Fig.\,2). The gray vertical bar denotes the I-M transition.
		 The inset shows the SdH oscillations in the KI state of YbB$_{12}$. {\bf d,} $T$ dependence of $\Delta f$ at
		 different fields. Data was taken at $\theta$ = 10.9$^\circ$ (Extended Data Fig.\,2b).
		 Solid lines are the Lifshitz-Kosevich (LK) fits. Inset: the field dependence of the fitted cyclotron energy $E_c^*$.}
		\label{fig:figure1}
	\end{center}
\end{figure}

\begin{figure}[hbtp]
	\begin{center}
		\includegraphics[width=1.0\linewidth]{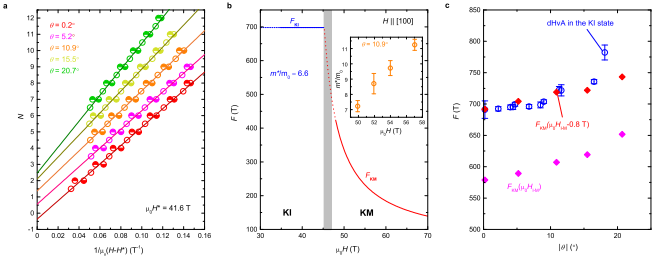}
		\caption{{\bf The field-dependent Fermi surface in the metallic state.}
		{\bf a,} The nonlinear Landau level plots shown in Fig.\,1c become linear after adding an offset of $\mu_0H$ = 41.6\,T
		to the applied magnetic field. The linear fits yield slopes which are denoted by the parameter $F_0$ in Eq.\,\ref{OrbitF}.
		The angular dependence of $F_0$ is summarized in Extended Data Table\,1. The Landau diagrams are shifted vertically by
		index $N$ = 1 for clarity. {\bf b,} With field applied along the [100] direction, the quantum oscillations in the KI state exhibits a
		field-independent frequency $F_{\rm KI}$, whereas in the KM state the SdH frequency $F_{\rm KM}$ is described by
		Eq.\,\ref{OrbitF}. Solid lines and short-dashed lines represent the field range in which SdH oscillations
		are detected and absent, respectively. The light gray vertical bar marks the I-M transition. Inset:
		The field dependence of the cyclotron masses $m^*$ inferred from the cyclotron energy $E_c^*$ shown in Fig.\,1d.
		{\bf c,} Angle dependence of quantum oscillation frequencies in YbB$_{12}$. Blue circles are dHvA frequencies in
		the KI state measured by torque magnetometry \cite{Xiang2018}.
		Magenta and red diamonds are SdH frequencies $F_{\rm KM}$ calculated using Eq.~\ref{OrbitF} using the transition field
		$\mu_0 H_{\textrm{I-M}}$ (Extended Data Table\,1) and $B$ = $\mu_0 H_{\textrm{I-M}}$-0.8\,T, respectively.  }
		\label{fig:figure2}
	\end{center}
\end{figure}

\begin{figure}[hbtp]
	\begin{center}
		\includegraphics[width=\linewidth]{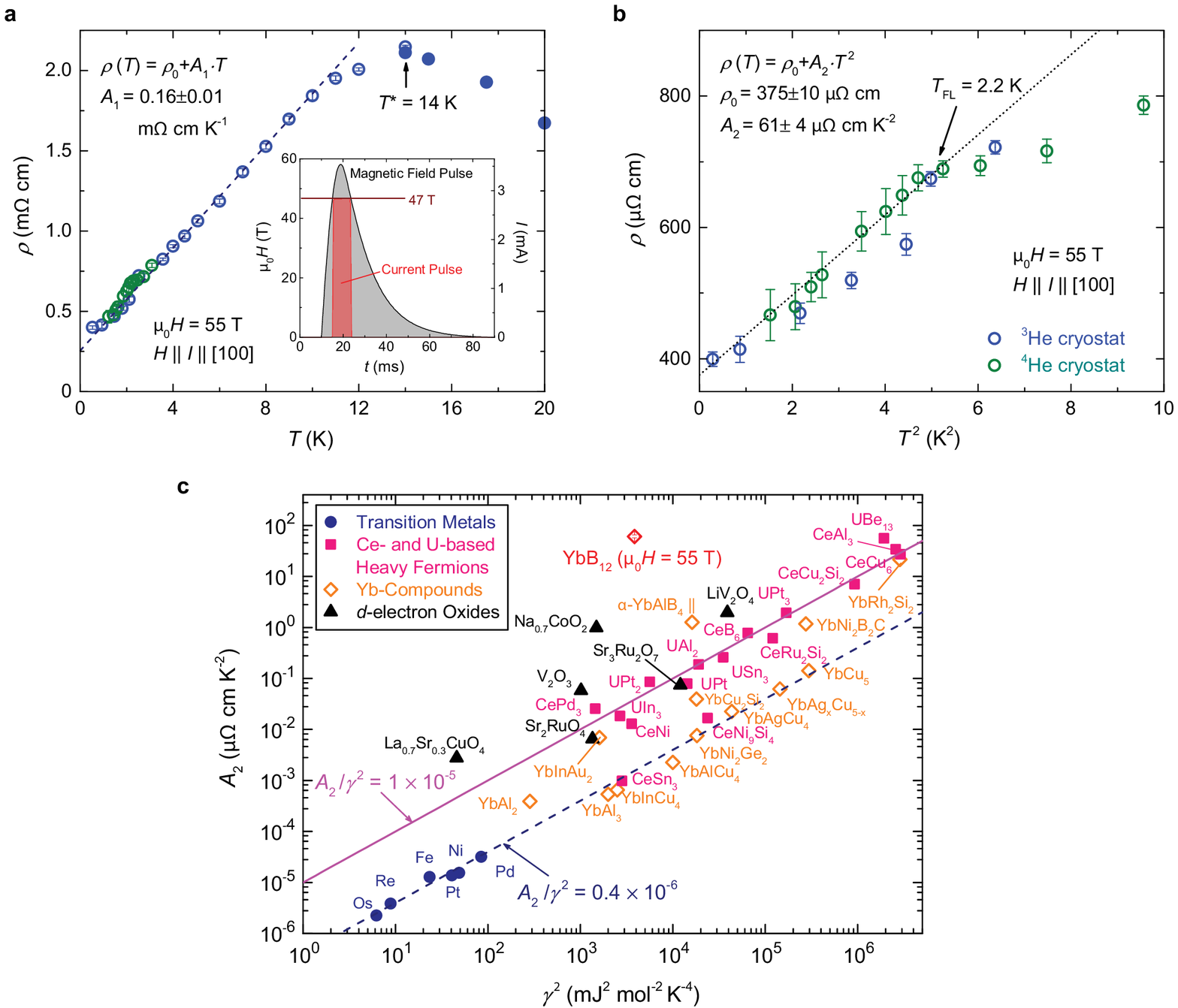}
		\caption{{\bf The temperature dependence of the resistivity in the metallic state.} {\bf a,} The resistivity of YbB$_{12}$ sample at
		$\mu_0 H$ = 55\,T plotted as a function of $T$. Both the current and the magnetic field were applied along the
		[100] direction. Blue and green hollow symbols are data measured using the pulsed current technique (see Methods) in
		$^3$He liquid or gas and $^4$He liquid, respectively. The solid symbols are data taken with a constant excitation in a
		$^3$He cryostat. A maximum in $\rho$ appears at $T^* = 14$~K. The dashed line is a linear fit of $\rho(T)$
		from 4\,K to 9\,K. The inset illustrates the magnetic field pulse and current pulse in our measurement in the time domain.
		{\bf b,} The same data plotted against $T^2$. The dotted line is a linear fit to the $^4$He-liquid data, showing the behaviour
		of the $T^2$-dependence below $T_{\textrm{FL}} = $2.2\,K.  {\bf c,} The deviation of YbB$_{12}$ from the
		Kadowaki-Woods relation. We use the value of the Sommerfeld coefficient $\gamma$ reported for YbB$_{12}$ in
		ref.\,\cite{TerashimaPRL2018}. The data points for transition metals, $d$-electron oxides and Ce- and U-based heavy fermions are taken
		from refs.\,\cite{Jacko,Hussey}, whereas the data for Yb-based compounds are taken from refs.\,\cite{Tsujii,Matsumoto}. }
		\label{fig:figure3}
	\end{center}
\end{figure}


\begin{thebibliography}{99}
\bibitem{gaugefield}
The linear Landau diagrams shown in Fig.\,2a with an ``offset field'' $H^*$
subtracted from the applied magnetic field bears a qualitative similarity to the composite-fermion
interpretation of the fractional quantum Hall effect (FQHE) observed in two-dimensional systems~\cite{Leadley}.
If we consider that this similarity is not coincidental, then $H^*$ can be regarded as an analogue of the
gauge field of the composite fermions in the FQHE.
In three-dimensional materials, an emergent gauge field, which can be felt by the charge carriers and
manifest itself in transverse transport measurements, might originate from chiral spin textures in a skyrmion
lattice phase \cite{Nagaosa}. However, such a topologically nontrivial magnetic structure has not yet been
proposed for YbB$_{12}$.
\end{thebibliography}
\end{document}